\documentclass[12pt]{iopart}

\usepackage{graphicx}
\usepackage{hyperref}
\usepackage{color}
\begin{document}

\title[From confinement to deconfinement in rectangular spin ices]{From confinement to deconfinement of magnetic monopoles in artificial rectangular spin ices }

\author{F.S. Nascimento, L.A.S. M\'{o}l, W.A. Moura-Melo and A.R. Pereira}

\address{Departamento de F\'isica, Universidade Federal de Vi\c cosa, Vi\c cosa, 36570-000, Minas Gerais, Brazil}
\eads{\mailto{fabiosantos.ba@gmail.com}, \mailto{lucasmol@ufv.br}, \mailto{winder@ufv.br}, \mailto{apereira@ufv.br}}

\submitto{\NJP}

\begin{abstract}
We study a frustrated two-dimensional array of dipoles forming an artificial rectangular spin ice with horizontal and vertical lattice parameters given by $a$ and $b$ respectively. We show that the ice regime could be stabilized by appropriate choices for the ratio $\gamma \equiv a/b$. Our results show that for $\gamma \approx \sqrt{3}$, i.e., when the center of the islands form a triangular lattice, the ground state becomes degenerate. Therefore, while the magnetic charges (monopoles) are excitations connected by an energetic string for general rectangular lattices (including the particular case of a square lattice), they are practically free to move for a special rectangular lattice with $\gamma \approx \sqrt{3}$. Besides that, our results show that
for $\gamma > \sqrt{3}$ the system is highly anisotropic in such a way that, even for this range out of the ice regime, the string tension almost vanishes along a particular direction of the array. We also discuss the ground state transition and some thermodynamic properties of the system.

\end{abstract}

\maketitle

%==========================================================================================
\section{Introduction}
%==========================================================================================

In the recent years, a great deal of efforts has been dedicated to the study of materials with frustrated interactions in an attempt to find and understand new states of matter \cite{Balents2010,Ramirez1999,Harris1997,Ryzhkin2005,Castelnovo2008,Wang2006,Mol2009,Mengoti2010,Ladak2010}. In particular, for ferromagnetic materials, two geometric structures are often investigated: the three-dimensional ($3d$) pyrochlore spin ice \cite{Ramirez1999,Harris1997,Ryzhkin2005,Castelnovo2008,Bramwell2009} and the two-dimensional ($2d$) analog thereof, the artificial kagome spin ice \cite{Mengoti2010,Ladak2010,Moller2009,Chern2011,Schumann2012,Branford2012}. These systems share the common phenomenon of a highly degenerate classical ground state. Besides, they also have excitations that are similar to Dirac magnetic monopoles \cite{Castelnovo2008,Mengoti2010,Dirac1931}. The artificial two-dimensional square spin ice is another structure very well studied, where monopoles excitations also appear \cite{Wang2006,Mol2010,Morgan2011,Morgan2011_2,Silva2012,Budrikis2012}; however, the square ice does not exhibit a degenerate ground state, since the two topologies that obey the ice rule have different energies \cite{Wang2006}. The main reason for this difference is the fact that, unlike the case of a tetrahedron in the natural $3d$ spin ice, the six bonds between the four islands belonging to a vertex are not all equivalent. As a first consequence, the square array provides a different type of magnetic monopole-like excitation: a kind of Nambu pair of monopole-antimonopole \cite{Mol2010,Silva2012,Nambu1974}, in which the opposite charges are effectively interacting by means of the usual Coulombic law plus a linear confining potential, the latter being related to a stringlike excitation binding the monopoles \cite{Mol2009,Mol2010}. Therefore, these monopoles are confined by a string similar to quark confinement in quantum chromodynamics \cite{Nambu1974}.

In a recent paper M\"{o}ller and Moessner \cite{Moller2006} showed that the problem related to the two inequivalent bond energies in the square ice could be remedied by introducing a height displacement $h$ between magnetic islands pointing in the horizontal and vertical directions. In this modified lattice, the ground state changes its configuration at a critical value of $h$ given by $h=h_{1}\approx 0.444b$ ($b$ is the lattice spacing) \cite{Mol2010,Moller2006}. Indeed, the string tension connecting the opposite monopoles decreases as $h$ increases from zero, almost vanishing at $h_{1}$. At this particular value of $h$, the ground state should become degenerate and the monopoles would become deconfined, similar to those found in the natural $3d$ spin ice compounds \cite{Castelnovo2008}. However, there is a price to be paid. The modified array is also three-dimensional and, in principle, it is considerable more difficult to be realized artificially. To our knowledge, such a system was not built yet.

%------------------------------------------------------------------------------------------
\begin{figure}[p]
	\centering
	\includegraphics[width=0.85\textwidth]{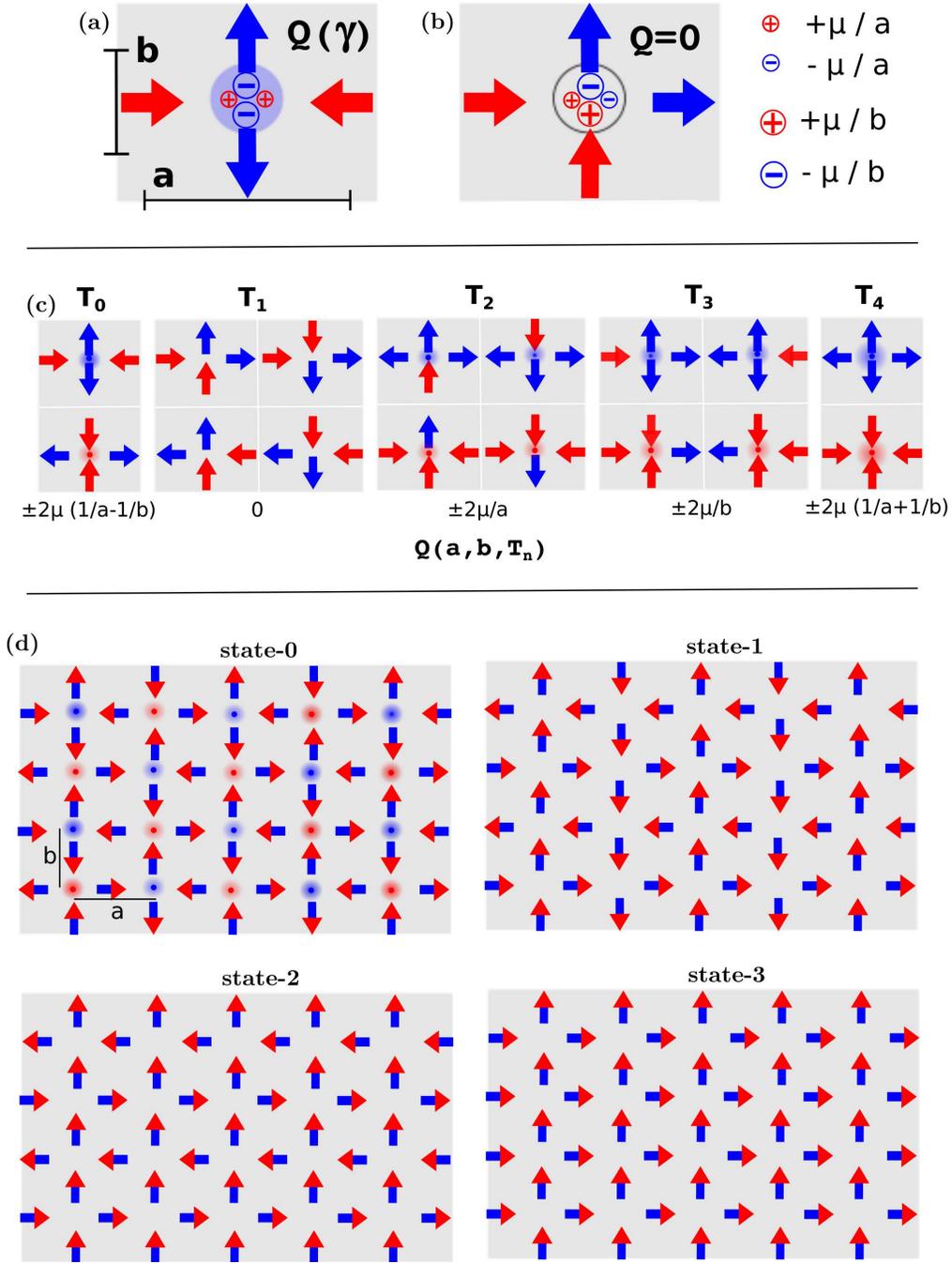}
	\caption{The rectangular lattice. Here, $a$ and $b$ are the horizontal and vertical lattice spacing, respectively. The arrows represent the spin orientation. In (a) and (b) we illustrate the $T_0$ and $T_1$ topologies and the residual charge due the dipole fractionalization. Part (c) shows the $16$ possible magnetic moment configurations on a vertex and the five distinct topologies, along with the magnetic charge obtained by the dumbbell picture. In (d) we show the configuration of a state denoted as state-0; the configuration of the dipoles looks like the ground state of the usual square spin ice. One can also see other configurations denoted as state-1, state-2 and state-3. The state-1 represents the ground state for $a/b > \sqrt{3}$. }
	\label{fig:rede}
\end{figure}
%------------------------------------------------------------------------------------------

On the other hand, there is a simpler way of dealing with this difficulty without transforming the array in a three-dimensional one. Indeed, one could mimic techniques used for natural systems in which the lattice spacing can even be tuned continuously by applying pressure along a particular direction \cite{Zhou2011}. Here, for a two-dimensional array, one does not need to apply any pressure to deform the lattice; it is sufficient to fabricate samples with an intentional change in either the horizontal or the vertical lattice spacing on the original square spin ice (``compressing" or ``stretching" the lattice, see Figs.\ref{fig:rede}(a)-(d)). Really, such a deformation can tune the ratios of the interactions between neighboring elements resulting in different magnetic ordering of the system as shown by Li \textit{at al.} \cite{Li2011}. Then, the ground state of this system is modified if the ratio of vertical to horizontal lattice spacing is $\sqrt{3}$ (or equivalently $1/\sqrt{3}$), which is equivalent to place the centers of the nanoislands in a triangular lattice. This critical value leads to a degenerate ground state, suggesting a residual entropy at absolute zero temperature similar to what happens to the natural and artificial $3d$ spin ice materials. In this paper we study the ground state, elementary excitations and thermodynamics of rectangular lattices which are characterized as a function of the assigned lattice anisotropy.

%==========================================================================================
\section{Model and Methods}
%==========================================================================================

At each site $(x_i,y_i)$ of the rectangular array (Fig. \ref{fig:rede}(d)), two spin variables are defined:
$\vec{S}_1 = \pm (1,0)$ located at $\vec{r}_1 = (a x_i + a/2, b y_i)$ and
$\vec{S}_2 = \pm (0,1)$ located at $\vec{r}_2 = (a x_i,b y_i + b/2)$.
The parameters $a$ and $b$ denote the horizontal and vertical lattice spacings, respectively (here, $b$ will be the standard lattice spacing, while $a$ is varied to give different rectangular arrays). When the parameters are equals ($a=b$) we then recover the square lattice \cite{Wang2006,Mol2009,Morgan2011,Silva2012}.
Therefore, in a lattice of area $A = l^2 b^2$ one gets $2 \times l^2 b / a$ spins. In this work we study the ratio $\gamma = a / b$ from $0.4$ to $4.0$ and we have fixed $b=1$. Representing the spins by $\vec{S}_i$, then the system is described by the following Hamiltonian
%------------------------------------------------------------------------------------
\begin{equation}
\label{eq:Hsi}
 H_{SI} = D  \sum_{i \neq j}\left[
 	\frac{\vec{S}_i \cdot \vec{S}_j}{r_{ij}^3}
 	- \frac{3 (\vec{S}_i \cdot \vec{r}_{ij})(\vec{S}_j \cdot \vec{r}_{ij})}{r_{ij}^5}
\right],
\end{equation}
%------------------------------------------------------------------------------------
where $D = \mu_0 \mu^2 / 4 \pi $ is the coupling constant of the dipolar interactions.

Each rectangular lattice vertex has $16$ possible magnetic moment configurations that can be characterized by five distinct topologies denoted by $T_{\nu}$($\nu =0,1,2,3,4$, see Fig. \ref{fig:rede}(c)). We remark that the square lattice admits only four different topologies \cite{Wang2006,Mol2009,Morgan2011}. Indeed, for the particular case in which $\gamma = 1$, topologies $T_2$ and $T_3$  have the same energy; on the other hand, for any $\gamma \neq 1$, $T_2$ and $T_3$  have different energies.

Another important difference between the cases $\gamma \neq 1$ and the square lattice ($\gamma =1$) is the emergence of residual magnetic charges even for topology $T_{0}$, which obeys the ice rule where two spins point inward and two spins point outward ($2-in, 2-out$) on a vertex. Although the dumbbell picture proposed in Ref.~\cite{Castelnovo2008} can not be simply transposed to the artificial square spin ice (since it does not describe the system quantitatively), it can help us to understand (qualitatively) some differences between the square and rectangular spin ices. The dumbbell picture~\cite{Castelnovo2008} is obtained when each spin in the lattice is replaced by a pair of magnetic charges placed on the adjacent vertices. Due to the lattice anisotropy, the horizontal spins should fix opposite magnetic charges (at adjacent vertices) given by $ q_h = \pm \mu/a$. On the other hand, the vertical spins would give adjacent opposite charges given by $ q_v = \pm \mu/b$. Therefore, at a vertex $(x_i,y_i)$ described by topology $T_{0}$, we can associate a residual charge with modulus $Q= \mid 2\mu/a - 2\mu/b \mid$ (see Fig.\ref{fig:rede}(a) and (c)). Of course, the residual charges present in topology $T_{0}$ vanish for the square lattice in which $a=b$. Note that topology $T_{1}$, which also obeys the ice rule, does not exhibit a residual charge (Fig.\ref{fig:rede}(b)); it has, however, a residual magnetic moment (here, referred to as ``SPIN"), not present in topology $T_{0}$. There is therefore an interesting asymmetry for the two topologies that obey the ice rule: on the vertices of $T_{0}$, one can find residual magnetic charges but not ``SPINS" while on the vertices of $T_{1}$ one finds residual ``SPINS" but the neutrality of magnetic charges is preserved. As $a/b$ is increased, the energy of the magnetic charge $Q$ (for $T_{0}$) increases while the energy of the effective ``SPIN" (for $T_{1}$) decreases. Such properties of the vertices obeying the ice rule (charge for $T_{0}$ or ``SPIN" for $T_{1}$) may play important roles in the ground state of the rectangular array. It may eventually exist a balance of the values of charge in $T_{0}$ and ``SPIN" in $T_{1}$ in such a way that the energies of these two topologies become equal for a particular $a/b$, causing a degenerate ground state. This possibility is studied in the next section. In Fig.~\ref{fig:rede}(c) the residual vertex charge $Q$ (in the dumbbell picture) is shown as a function of the lattice parameters $a$ and $b$. An interesting point is that topologies $T_2$ and $T_3$ are degenerate in the square lattice ($\gamma=1$) and can be grouped in a single topology. However, as long as $\gamma \ne 1$ they have different energies and charges. For instance, considering $\gamma > 1$ ($a>b$), vertices on topology $T_2$ has more charge than vertices on topology $T_3$. This difference has important consequences on charges interactions as will be shown soon. Of course, as stated above, the dumbbell picture can not be used to quantitatively describe artificial spin ices such that we do not expect that the charges dependence on lattice spacings $a$ and $b$ are indeed those shown in Fig.~\ref{fig:rede}(c).

%==========================================================================================
\section{Ground states}
%==========================================================================================
%------------------------------------------------------------------------------------------
\begin{figure}[b]
	\centering
	\textbf{\scriptsize(a)}\\ \includegraphics[width=8 cm]{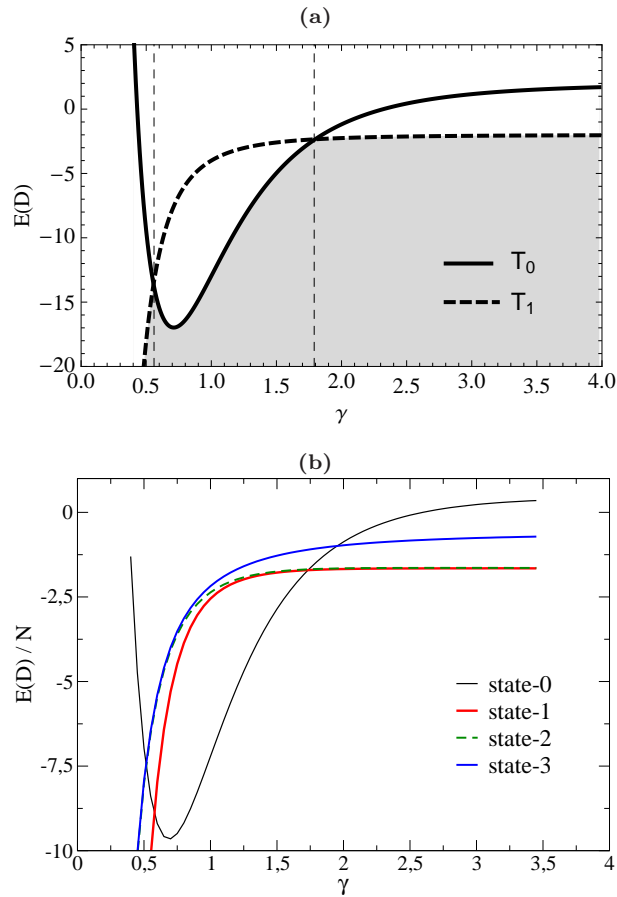} \\
	\textbf{\scriptsize(b)}\\ \includegraphics[width=8 cm, clip=true]{fig2b.eps}
	\caption{(a) Analytical result for the energy of a vertex in topology $T_0$ and of a vertex in topology $T_1$ as a function of the parameter $\gamma$. (b) Energies of a lattice with 800 spins in the state-0 (solid black curve), state-1 (solid black curve), state-2 (dashed green curve) and state-3 (solid blue curve) as a function of the parameter $\gamma$.}
	\label{fig:energiaVSdx}
\end{figure}
%------------------------------------------------------------------------------------------

It is well known that, on a square lattice ($\gamma=1$), the topology $T_0$ is more energetically favorable than the others. Consequently, the ground state of an artificial square spin ice has a configuration with all vertices obeying the ice rule with topology $T_{0}$ as illustrated in Fig. \ref{fig:rede}(d) state-0. In this case the ground state is only twofold degenerate. It is also valid for an appreciable range of values ($\gamma \neq 1$) of the rectangular ice. Nevertheless, the energy of topologies $T_0$ and $T_1$ approaches each other when we set the dipoles on a rectangular geometry such that $\gamma \to 0.556$ or $\gamma \to 1.797$ (see Fig. \ref{fig:energiaVSdx}$(a)$). Indeed, the analytical expression for the energy difference ($\Delta E$) between a vertex on topology $T_0$ and a vertex on topology $T_1$ as a function of $\gamma$ is

$$\Delta E=4\left(1+\frac{1}{\gamma^3}-\frac{24\gamma}{(1+\gamma^2)^{5/2}}\right),$$
%---------------------------------------------------------------------------------
which gives the above numerical results. Then, there are two special ratios $\gamma$ that may degenerate the ground state of the system, since the topologies satisfying the ice rule have the same energy for $\gamma = 0.556$ and $\gamma = 1.797$. However, the values of $\gamma$ in which the system is degenerate may not be those obtained in a single vertex analysis, since long range interactions are present. To get the correct values we have evaluated the energy (see Eq.~\ref{eq:Hsi}) for the four different ordered configurations of spins as shown in Fig.~\ref{fig:rede}(d) using different lattice sizes. A typical curve, obtained for a lattice with $800$ spins, is shown in Fig.~\ref{fig:energiaVSdx}(b). From a finite size scaling analysis we see that the values are $\gamma_{c1} =1.73287(5)\approx \sqrt{3}$ and $\gamma_{c2}=0.57707(2)\approx 1/\sqrt{3}$.
Moreover, for $\gamma_{c1} < \gamma < \gamma_{c2}$, the ground state is composed by topology $T_0$ (possessing residual charges alternating the signs at adjacent vertices) and for $\gamma < \gamma_{c1}$ or $\gamma > \gamma_{c2}$, the ground state is dominated by topology $T_1$ (possessing residual ``SPINS" located at the vertices). There are then two equivalent ground state transitions on a rectangular lattice.
State-$0$ (see Fig.~\ref{fig:rede}(d)) is the ground state of a square array (and also of a rectangular array for $\gamma_{c1} < \gamma < \gamma_{c2}$) with all vertices described by topology $T_0$; State-$1$ is the ground state of a rectangular array with $\gamma < \gamma_{c1}$ or $\gamma > \gamma_{c2}$ with all vertices described by topology $T_1$. There are other configurations that also obey the ice rule with topology $T_1$ but they are not ground states since they have higher energies than that of state-$1$. Note also that state-$1$ has null total magnetization while state-2 is magnetized along the ``vertical'' direction and state-$3$ is magnetized along the diagonal. Therefore, the ground state of the rectangular lattice is either, a state with residual charge at each vertex but with null net charge or a state with residual spins at each vertex but with null net magnetization.

A simple argument can lead to the above approximate quantities: by replacing the net magnetic moment of the islands by a point-like dipole at their centers, one can easily see that, for any vertex of a square array, the distance $b$ between the two adjacent spins pointing along the same direction is smaller than the distance $\sqrt{2}b$ between the two adjacent spins pointing along perpendicular directions. However, if the lattice is deformed toward the rectangular shape, the lines connecting the spins pointing along perpendicular directions form a rhombus with a short diagonal of length $b$ and a long diagonal of length $a$. For the special ratios $\gamma =\sqrt{3}$ or $\gamma =1/\sqrt{3}$, two back to back equilateral triangles are formed, in such a way that the distance between spins pointing along different directions (the length of the sides of the rhombus) becomes equal to the distance between one of the two pairs of spins pointing along the same direction (the length of the rhombus' short diagonal). Then, for this special case, the energies of the different spins configurations become degenerate. The square spin ice was thus transposed to a triangular lattice.

As we have seen, the ground state configuration on rectangular lattices has a dependence on the parameter $\gamma$. However, due to rotational symmetry of the Hamiltonian, the process of compressing or stretching the lattice is equivalent to a deformation. Indeed, the parameters $\gamma_{c1}$ and $\gamma_{c2}$ are just the reverse of each other, since it is enough to rotate the sample by an angle $\pi/2$ to pass from $\gamma$ to $1/\gamma$. For this reason we shall consider from now on only the stretching process, i.e., $\gamma \geq 1$. Thus, for $1 < \gamma < \gamma_{c2}$, the ground state of a rectangular array assumes the same configuration of the square spin ice ground state (state-$0$), as illustrated in the Fig. \ref{fig:rede}$(d)$. In particular, for this range, the ground state exhibits an interesting ordering of alternating positive and negative charges $Q$ on the vertices. In addition, such residual charges are energetically favorable than residual ``SPINS" found on topology $T_{1}$.

On the other hand, for $\gamma > \gamma_{c2}$, the lattice anisotropy forces the system to assume a modified ground state (state-$1$) which is also illustrated in Fig. \ref{fig:rede}$(d)$. For this range, the residual ``SPINS" are more energetically favorable than the residual charges of topology $T_{0}$. The balance between local residual charges (found in topology $T_{0}$) and local residual ``SPINS" (found in topology $T_{1}$) must be eventually established (from the energetic point of view) when $\gamma = \gamma_{c2}$, making these two topologies to become degenerate.

%-----------------------------------------------------------------------------------
\begin{figure}[t]
	\centering
	\textbf{\scriptsize(a)}~\includegraphics[width=7.5 cm, clip=true]{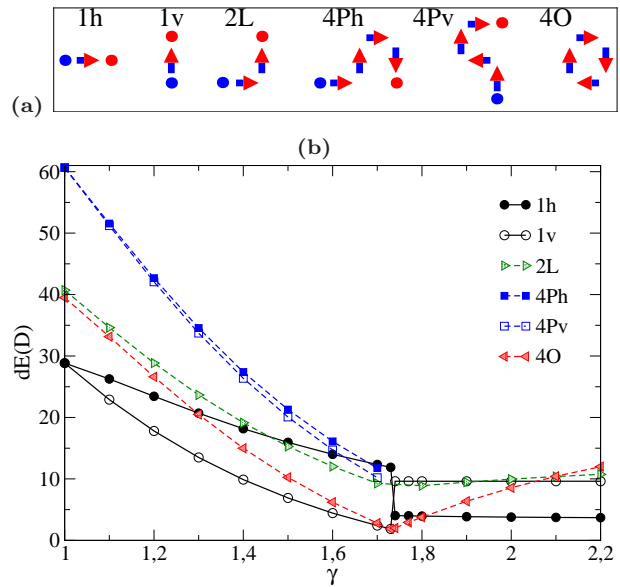}\\
	\textbf{\scriptsize(b)} \\ \includegraphics[width=8 cm, clip=true]{fig3b.eps}
	\caption{(a) Some low energy excitations and their mnemonic symbol. (b) Energies of these excitations as a function of the parameter $\gamma$.}
	\label{fig:excitacaoVSdx}
\end{figure}
%-----------------------------------------------------------------------------------

%==========================================================================================
\section{Excitations and monopole deconfinement}
%==========================================================================================
The inversion of a single spin (for example, between adjacent vertices $i$ and $i+1$) violates the ice rule generating an excited state, which implies in an excess of opposite magnetic charges placed on $i$ and $i+1$. It is a pair of defects similar to monopole quasi-particles, positioned at two adjacent vertices. It is useful here to distinguish the different types of monopole defects: the most energetic ones, in which the spins are in the ($4-in$) or ($4-out$) states ($T_4$ in the Fig. \ref{fig:rede}(c)) and the lower energetic ones, in which the spins are in the ($3-in,1-out$) or ($3-out,1-in$) states (topologies $T_2$ and $T_3$ shown in Fig. \ref{fig:rede}(c)). It is important to remark that topologies $T_2$ and $T_3$ have different charges and energies for general $\gamma \neq 1$ (see Fig.~ \ref{fig:rede}(c)). Therefore, the system may be filled by several elementary excitations that can be classified by the number of flipped moments and a mnemonic character for shape as introduced in Ref. \cite{Morgan2011}. However, here, there is a small difference that introduces a new factor in the classification: separations of the magnetic charges along the horizontal ($h$) and vertical ($v$) lines of the array have different energies. As examples, the simplest excitations are symbolized by $1v$ and $1h$, representing two opposite charges separated by one lattice spacing along the vertical and horizontal lines respectively (excitation $1v$ creates two adjacent vertices in topology $T_2$ and excitation $1h$ creates two adjacent vertices in topology $T_3$). The separation process of the monopoles generates energetic one-dimensional strings of dipoles that can be seen as lines which pass by adjacent vertices that obey the local ice rule. In Fig. \ref{fig:excitacaoVSdx} we present the symbol of some pairs of monopoles and their energy cost as a function of $\gamma$. The behavior of the energy cost as a function of $\gamma$ is not trivial, but, in general, for $\gamma < \gamma_{c2}$ we observe that the energy cost of the defect decreases as the lattice is stretched. Note that we did not show the energies of the excitations $4Ph$ e $4Pv$ for $\gamma > \gamma_{c2}$ since these excitations generate extra charges on the ground state. One might naively think that the energetic cost of these excitations should follow the dependence on lattice spacings shown in Fig.~\ref{fig:rede}(c). However, as said before, the dumbbell picture does not quantitatively describe the system and thus, the dependence on $\gamma$ is not really that of Fig.~\ref{fig:rede}(c). Indeed, we could not find a simple law relating the energetic cost with the magnetic charge of the excitations (which will be considered latter).

%-----------------------------------------------------------------------------------
\begin{figure}[b]
	\centering
	\includegraphics[width=8.0 cm, clip=true]{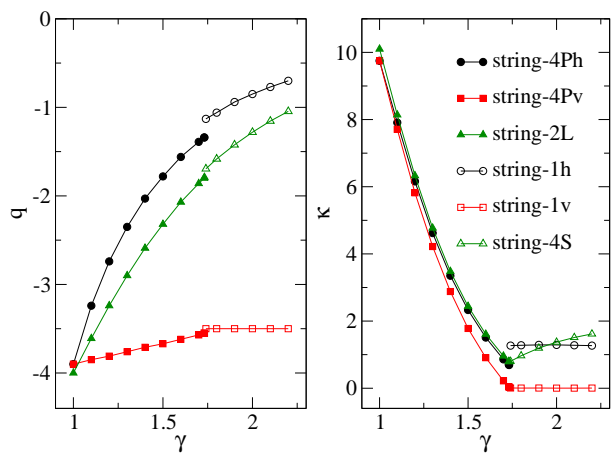}
	\caption{The parameters $q$ (Left) and $\kappa$ (Right) as a function of $\gamma$. The circle, square and triangle are the results when the monopoles are separated along the directions horizontal, vertical and diagonal, respectively.}
	\label{fig:tensaoVSgamma}
\end{figure}
%-----------------------------------------------------------------------------------

 In recent works \cite{Mol2009,Mol2010} considering the square array, we have proposed that the most general expression for the total cost of a monopole-antimonopole pair separated by a distance $R$ is the sum of the usual Coulombic term roughly equal to $q/R$, and a term roughly equal to $\kappa X$ resulting from the string joining the monopoles (here, $X$ is the string length; there is still a constant term associated with the pair creation energy \cite{Mol2009,Mol2010}). The parameters $q$ and $\kappa$ have a small dependence on the direction in which the monopoles are separated in the crystal plane \cite{Mol2010}. This anisotropy is therefore manifested in both the Coulomb and linear terms of the potential. Although the monopoles interaction should be highly dependent of stretching of the lattice, one should expect that, for the rectangular lattice, in general, things must have some similarities with the square lattice.
 The main difference is that the square array has only one kind of unit-charged monopole while the rectangular array has two (topologies $T_2$ and $T_3$). However, the quantity that measures the interaction between monopoles involves the product between these charges, i.e., the quantity $q$ of Refs.~\cite{Mol2009,Mol2010} should be somewhat like $q=q_1 q_2$ where $q_1$ and $q_2$ are the charges of each monopole (the remaining term, associated with the string tension, is supposed to be independent of the charges itself). So, as long as we can always deal with the same kind of charges (considering for example only the interactions between vertices of the same kind) or even between vertices of different types, we expect that the same phenomenology of the square spin ice applies to the artificial rectangular spin ice. Indeed, the function
%
%
%--------------------------------------------------------------------------------------
\begin{equation}
\label{eq:potential}
 V(R) = \frac{q(\phi,\gamma)}{R} + \kappa(\phi,\gamma)X + c(\phi,\gamma),
\end{equation}
%--------------------------------------------------------------------------------------
is expected to describe the interactions between the monopoles in both cases (here, $\phi$ is the angle that the line joining the monopole-like defects makes with the $x$-axes of the array, and $q$ is the product between monopoles charges).

Before showing the results, a detailed discussion on how they were obtained would be useful. First we would like to remark that the potential $V(R)$ is simply the energy of each configuration of spins minus the ground state energy. Each point of this curve is obtained by evaluating the system's energy for a given configuration for which the distance between the monopoles is $R$. One very important point of our calculations is that we must establish a connection between the string length ($X$) and the distance between the monopoles ($R$), otherwise we would have two variables in our model. This is why we have chosen, in Refs.~\cite{Mol2009,Mol2010}, two different shapes ($2L$ and $4P$) as the building blocks of the string, such that $X=\sqrt{2}R$ for the $2L$ strings and $X=2R$ for $4P$ strings. Note also that we have to keep the neutrality in the region between the two opposite monopoles (in the context of two-in/two-out, while flipping spins in sequence); then, after flipping an horizontal spin, a vertical one must be flipped and so on. In this context, using the same string shapes of Refs.~\cite{Mol2009,Mol2010} ($2L$ and $4P$), we always have at the end points of a string, a type $T_2$ vertex and a type $T_3$ vertex for excitations above state-0. So, in this case, the constant $q$ is the product of different types of monopoles ($q=q_{T_2}q_{T_3}$). We remark that in the separation process, the moving monopole changes the topology (from $T_2$ to $T_3$ and vice-versa) in such a way that during this procedure, there will be interactions between the same kind of vertices ($T_2$ with $T_2$ for example) alternating with interactions between different kinds of vertices; however, we only keep in the expression of the potential used to fit Eq.~\ref{eq:potential}, the interactions between different kinds of vertices (since this is the only way of establishing a precise relationship between the string length $X$ and the distance between the monopoles $R$). One extra ingredient about the rectangular lattice is the fact that the separation of monopoles along the $x$ or $y$ direction is different, such that we have introduced the notation $4Ph$ and $4Pv$ strings. Besides that, the string in the rectangular lattice is composed by $T_1$ vertices, which has neutral charge. However, since the ground state (with the configuration of state-0) is a charge's crystal, the string itself must have a net magnetic charge. Indeed, this net charge of the string is responsible for keeping the system's neutrality, once $T_2$ and $T_3$ vertices have different charges, in such a way that the composite object, string plus monopoles, is neutral. For excitations above state-1, things are less complicated since the monopoles can be separated along a straight line by a sequence of type $T_1$ vertices of opposite ``SPIN'', which are neutral in the ground state. In this way, only one kind of monopole can be considered (interaction between either $T_2$ or $T_3$ vertices) and the string length $X$ is equal to the distance between monopoles. Of course, it is also possible to chose other shapes for the string and separate the charges along other directions. In this case, straight portions of the string are composed by $T_1$ vertices, while, along the region it bends (making a corner) the topology of the string changes to $T_0$ and thus the monopole changes from topology $T_2$ to $T_3$ or vice-versa. Indeed, we have also considered strings where the building block is a $4S$ excitation (formed by joining an $1h$ with two consecutive $1v$ and another $1h$ excitations) as shown in Fig. 5 picture (3) of Ref.~\cite{Mol2010}.

Figure \ref{fig:tensaoVSgamma} shows the dependence of the parameters $q$ and $\kappa$ on $\gamma$. The parameter $q$ for $1 < \gamma < \gamma_{c2}$ is $q=q_{T_2}q_{T_3}$ since, as discussed above, the monopoles generated by the string shapes $4Ph$, $4Pv$ and $2L$ are described by topologies $T_2$ and $T_3$. This quantity has a clear dependence on $\gamma$ (the expected dependence from the dumbbell picture is $1/\gamma$ as one can see in Fig~\ref{fig:rede}(c)); however, for each string shape, a different dependence is observed. This feature is not completely understood at the moment. The string tension is almost the same for these three string shapes in this region and diminishes as $\gamma$ increases, showing only a small anisotropy that increases with increasing $\gamma$. For $\gamma > \gamma_{c2}$, the string is a sequence of adjacent vertices with topology $T_1$, with its local ``SPINS'' in  different directions as discussed above. In this case, the dumbbell picture describes well the charge's dependence on $\gamma$. Indeed, for  $1h$ and $4S$ strings, both monopoles are in topology $T_2$ and thus one may expect a $1/\gamma^2$ dependence, as it really is. For string $1v$ both monopoles are in topology $T_3$ and no dependence on $\gamma$ should be expected as shown. The string tension is constant in this region for strings $1h$ and $1v$ and it increases with increasing $\gamma$ for $4S$ string, such that the system is anisotropic.
Then, for a rectangular array with $\gamma \geq \gamma_{c2}$, the monopoles become deconfined for separation along the vertical direction while they still remain  a bit confined for separation along the horizontal direction (there is a small but finite string tension along this direction; see Fig.\ref{fig:tensaoVSgamma}).

%-----------------------------------------------------------------------------------
\begin{figure}[b]
	\centering
	\textbf{\scriptsize(a)}\\\includegraphics[width=7 cm, clip=true]{fig5a.eps} \\
	\textbf{\scriptsize(b)}\\\includegraphics[width=7.5 cm, clip=true]{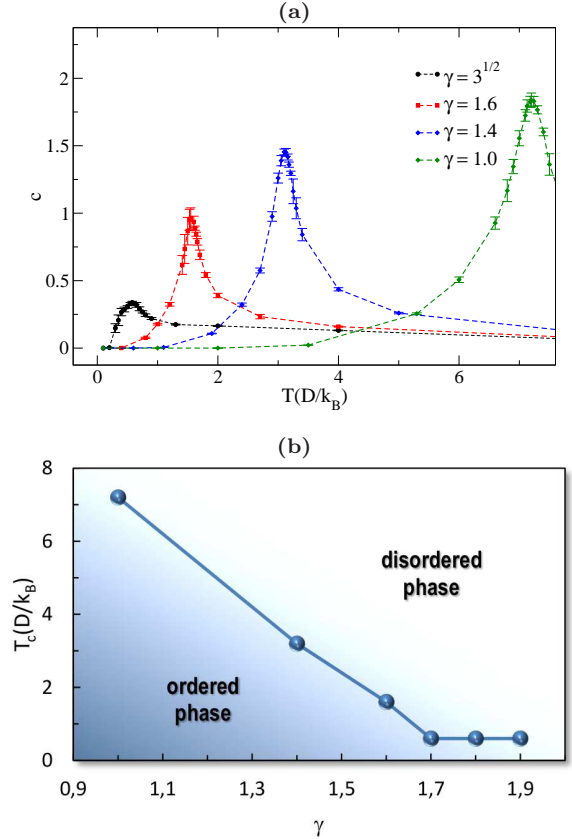}
	\caption{(a) Specific heat as function of the temperature for different values of $\gamma$. (b) Critical temperature as a function of $\gamma$.}
	\label{fig:cT}
\end{figure}
%-----------------------------------------------------------------------------------

%==========================================================================================
\section{Thermodynamics}
%==========================================================================================

Here, we perform standard Monte Carlo techniques to obtain the thermodynamics averages of the system defined by Hamiltonian \ref{eq:Hsi}. Periodic boundary conditions were implemented and our Monte Carlo procedure includes a combination of single spin flips and loop moves~\cite{Silva2012,Barkema1998}, where all spins contained in a closed random loop are flipped according to the Metropolis prescription. In our scheme, one Monte Carlo step (MCS) consists of $2 \times l^2 / \gamma$ single spin flips and $1$ loop move. Usually, $10^4$ MCS were shown to be sufficient to reach equilibrium configuration and we have used $10^4 - 10^6$ MCS to get thermodynamic averages. All results shown here were obtained for $l=20$.

Figure \ref{fig:cT}(a) shows the specific heat as a function of the temperature and $\gamma$. The sharp peak in the specific heat suggest that the system undergoes a phase transition as shown in Ref. \cite{Silva2012} for the square lattice ($\gamma=1$). Indeed, the height of these peaks increases logarithmically with the system size \cite{Silva2012}. The transition seems to be characterized by the appearance of several excitations (Nambu monopoles and string loops). Particularly, the monopoles may become free at $k_B T \approx 7.2 D $ as discussed at Ref.~\cite{Silva2012}. On the other hand, we can see that the lattice anisotropy changes the temperature of the transition and the peak in the specific heat has different positions and intensities, depending on the parameter $\gamma$. The critical temperature decreases as $\gamma$ increases from $1$ to $\gamma_{c}$ (see Fig. \ref{fig:cT}(b)). The fact that the string tension also decreases as $\gamma$ increases, practically vanishing at $\gamma=\sqrt{3}$, corroborates the argument of Refs. \cite{Mol2009,Silva2012} that the phase transition argued for the square arrays is associate with the string configurational entropy: for $X$ sufficiently large, the number of configurations of strings connecting two opposite monopoles would be well approximated by the random walk result  $p^{X/b}$ (for a $2d$ square lattice, $p = 3$). Then, using a very simple estimate, the string configurational entropy   ($ k_{B} ln [p^{X/b}]$) is proportional to $X$ \cite{Silva2012}, and the string free energy  can be approximated by $F=[\kappa-(\ln3)k_{B}T/b]X$, which leads to an estimate of the critical temperature as $T_{c} \sim b\kappa/ln(3) $. Therefore, $T_{c}\propto \kappa$ and, as one can observe in Figs.~\ref{fig:tensaoVSgamma} and \ref{fig:cT}, their dependence on $\gamma$ is very similar.

%==========================================================================================
\section{Conclusions}
%==========================================================================================
In summary we have studied a possible artificial spin ice array in two dimensions with rectangular geometry.
We have carried out the investigation of non-thermally activated order-disorder transitions on these possible artificial spin ice structures. By varying the ratio between vertical and horizontal lattice spacings ($a/b$) we have obtained several properties of the system, which may depend on the competition between residual charges and residual ``SPINS" in the ground states.  Such rectangular arrangement also contains elementary excitations which are similar to Nambu monopoles \cite{Nambu1974}: opposite charges confined by energetic strings like quarks in quantum chromodynamics. However, the special values $a/b=\sqrt{3}$ and $a/b=1/\sqrt{3}$ (which corresponds to place the islands in a triangular lattice) approach the array to the ice regime, increasing the frustration of the system which allows greater mobility for the monopoles. Then, in contrast to the  square lattice \cite{Mol2009,Mol2010}, monopoles may become free to move, even for very low temperatures, in a rectangular array with $a/b=\sqrt{3}$ or $a/b=1/\sqrt{3}$, leading to the possibility of monopole controls, opening an avenue for new technologies. So, here we have shown that there is a possibility of the existence of deconfined poles in two-dimensional artificial ices not due to thermal effects, but due to the lattice construction itself. Moreover, we have investigated some aspects of the thermodynamic of this system, which corroborate the idea of a phase transition induced by the string configurational entropy ~\cite{Silva2012}. In addition, we have shown that this particular change in the system's geometry can be interpreted in terms of residual magnetic charges at each vertex. Indeed, this picture is similar to that observed for a square spin ice where one of the islands is deformed, being smaller or greater than the others~\cite{Silva2012b}. This may indicate that the properties of an imperfect system may be modeled by the presence of residual charges on the vertices. Besides, we remark that changes in the system's geometry may be a simple and effective way in the search for artificial spin ice systems where monopoles excitations can be controlled.
Indeed, the dependence of the excitation's energies and consequently of the constants $q(\phi,\gamma)$, $\kappa(\phi,\gamma)$ and $c(\phi,\gamma)$ on $\gamma$ (see Figs.~\ref{fig:excitacaoVSdx} and \ref{fig:tensaoVSgamma})  may change the system's dynamics for different aspect ratios. For instance, the strength of the system's anisotropy depends on $\gamma$ and this fact may facilitate the dynamic process that drives the system to its ground state in the presence of external magnetic fields (for more details concerning the system dynamics in the presence of external fields see Refs.~\cite{Budrikis2010,Budrikis2011,Budrikis2012b}).
Finally, we would like to remark that an important fact to stabilize the ice regime is the equidistance between spins around the vertex centers. Indeed, one can also see it in other systems such as the tetrahedron of natural spin ices, in the artificial kagome and modified square spin ices \cite{Moller2006,Moller2009,Mol2010} and now, as more one example, in the artificial rectangular ice with special ratio $a/b=\sqrt{3}$.

\ack{The authors thank CNPq, FAPEMIG and CAPES for financial support.}

\section*{References}

%==========================================================================================
\bibliographystyle{unsrturl}
\bibliography{retangular}
%==========================================================================================

%///////////////////////////////////////////////////////////
			\end{document}